\documentclass[twocolumn,aps,pre,showpacs]{revtex4}
\usepackage[dvips]{graphicx}
\usepackage[usenames]{color}
\usepackage{amssymb,amsfonts,amsmath}
\usepackage{epsfig}
\begin{document}

\title{Optimal estimates of the diffusion coefficient of a single
Brownian trajectory}
%Distribution of the maximum likelihood estimator
%of a single Brownian trajectory diffusion coefficient

\author{Denis Boyer$^{1}$\footnote{boyer@fisica.unam.mx},
David S. Dean$^{2}$\footnote{david.dean@u-bordeaux1.fr},
Carlos Mej\'{\i}a-Monasterio$^{3}$\footnote{c.mejiamonasterio@gmail.com} and
Gleb Oshanin$^{4}$\footnote{gleb.oshanin@gmail.com}}

\affiliation{
$\ ^1$Instituto de F\'{\i}sica, Universidad Nacional Autonoma de M\'exico,
D.F. 04510, M\'exico\\
$\ ^{2}$Universit\'e de  Bordeaux and CNRS, Laboratoire Ondes et
Mati\`ere d'Aquitaine (LOMA), UMR 5798, F-33400 Talence, France\\
$\ ^{3}$Laboratory of Physical Properties, Technical University
of Madrid, Av. Complutense s/n 28040, Madrid, Spain\\
$\ ^{4}$Laboratoire de Physique Th\'eorique de la Mati\`ere
Condens\'ee (UMR CNRS 7600), Universit\'e Pierre et Marie Curie, 4
place Jussieu, 75252 Paris Cedex 5 France}

\date{\today}

\begin{abstract}
Modern developments in microscopy and image processing  are revolutionizing areas of physics, chemistry and biology as nanoscale objects can be tracked with unprecedented accuracy. The goal of single particle tracking is to determine the interaction between the particle and its environment.  The price paid for having a direct visualization of a single particle is a consequent lack of statistics. Here we
address the optimal way of extracting diffusion constants from single trajectories for pure Brownian  motion.  It is shown that the maximum likelihood estimator is much more efficient than the commonly used least squares estimate. Furthermore we investigate the effect of disorder on the distribution of estimated diffusion constants and show that it increases the probability of observing estimates much smaller than the true (average) value.
\end{abstract}

\pacs{05.40.Jc, 31.15.xk, 87.16.dp, 61.43.Er} \maketitle

\section{Introduction}

Single particle tracking dates back to the classic
study of Perrin on Brownian motion (BM) \cite{perrin}. It generates the
position time series of an individual particle trajectory ${\bf B}(t)$
in a medium (see, e.g., Refs.~\cite{bra,saxton}) and when properly interpreted,
the information drawn from a single, or a finite number of trajectories,
can provide insight into the mechanisms and forces that drive or
constrain the motion of the particle. The method is thus potentially a
powerful tool to probe physical and biological processes at the level
of a single molecule \cite{moerner}. At the present time, single
particle tracking is widely used to characterize the microscopic
rheological properties of complex media \cite{mason}, and to probe the active
motion of biomolecular motors \cite{greenleaf}. In biological cells
and complex fluids, single particle trajectory (SPT) methods have, in
particular, become instrumental in demonstrating deviations from normal
BM of passively moving particles (see, e.g.,
Refs.~\cite{golding,weber,bronstein,seisenberger}).

The reliability of the information drawn from  SPT analysis,
obtained at high temporal and spatial resolution but at
expense of statistical sample size is not always clear.  Time averaged
quantities associated with a given trajectory may be subject to large
fluctuations among trajectories. For a wide class of anomalous
diffusions described by continuous-time random walks, time-averages of
certain particle's observables are, by their very nature, themselves
random variables distinct from their ensemble averages
\cite{rebenshtok}.  An example is the square displacement
time-averaged along a given trajectory, which differs from the
ensemble averaged mean squared displacement \cite{he}. By analyzing
time-averaged displacements of a particular trajectory realization,
subdiffusive motion can actually look normal, although with strongly
differing diffusion coefficients from one trajectory to another
\cite{lubelski}.

Standard BM is a much simpler and exceedingly
well-studied random process \cite{yuval} than anomalous diffusion, but
still it is far of being as straightforward as one might be tempted to
think. Even in bounded systems, despite the fact that the first
passage time distribution has all moments, first passages to a given
target of two independent identical BMs, starting at the same point in
space, will most likely occur at two distinctly different time moments
\cite{carlos}, revealing a substantial manifestation of
sample-to-sample fluctuations.  Ergodicity, that is, equivalence of
time- and ensemble-averages of square displacement holds only in the
infinite sample size limit.  In practice, it means that standard
fitting procedures applied to finite (albeit very long) trajectories
of a given particle will unavoidably lead to fluctuating estimates of
the diffusion coefficient $D$.
Indeed, variations by orders of magnitude have been observed in SPT
measurements of the diffusion coefficient of
the LacI repressor protein along elongated DNA \cite{austin} (see also Section \ref{z}). 
Significant sample-to-sample fluctuations resulting in broad histograms for the value of the diffusion coefficient
have been observed experimentally for 
two-dimensional (2D) diffusion in the 
plasma membrane \cite{saxton}, as well
as for 
 diffusion of a single protein in the cytoplasm and nucleoplasm of
mammalian cells \cite{goulian}.

Such a broad dispersion of the value of the diffusion coefficient
extracted from SPT measurements, raises important
questions about the correct or optimal methodology that should be used
to estimate $D$. Indeed, these measurements are performed in rather
complex environments and each SPT has its own history
of encounters with other species, defects, impurities, etc., which
inevitably results in rather broad histograms for observed $D$.  On
the other hand, it is highly desirable to have a reliable estimator of
the diffusion coefficient even for the hypothetical "pure" cases, such
as, e.g., unconstrained standard BM. 
A reliable estimator should produce
a distribution of $D$ as narrow as possible 
and with the most probable value as close as
possible to the ensemble average one. 
A knowledge of the distribution
of such an estimator could provide a useful gauge to identify effects
of the medium complexity as opposed to variations in the underlying
thermal noise driving microscopic diffusion.  Commonly used methods of
extraction of $D$ from the SPT data are based on
a least square (LS) estimate of the time-averaged square displacement
and some of its derivatives (see, e.g., \cite{saxton,goulian,saxton2}
and the next section).  A recent study, Ref.~\cite{boyer}, focussed on
estimators for $D$ for 1D BM, the statistics of which is amenable to
analytical analysis.  Several methods for estimating $D$ from the SPT
 data were studied and it was shown that a completely
different approach - consisting of maximizing the unconditional
probability of observing the whole trajectory - is superior to those
based on the LS minimization.  As a matter of fact, at least in 1D
systems the distribution of the maximum likelihood (ML) estimator of
the diffusion coefficient not only appears narrower than the LS
ones, resulting in a smaller dispersion, but also the most probable
value of the diffusion coefficient appears closer to the ensemble
average $D$ \cite{boyer}.

In this paper we focus first on the case of pure standard BM and
calculate exactly, for arbitrary spatial dimension $d$, the
distribution $P(u)$ of the maximum likelihood estimator
\begin{equation}
\label{u}
u = \frac{1}{T} \int^{t_0 + T}_{t_0} dt \frac{{\bf B}^2(t)}{\mathbb{E}\left\{{\bf B}^2(t)\right\}},
\end{equation}
of the diffusion coefficient of a single BM trajectory ${\bf
  B}(t)$. The parameter $t_0$ here is the lag time (at which the
measurement is started) which can be set equal to zero for standard BM
without any lack of generality. However for anomalous diffusion, or BM
in presence of disorder $t_0$ will play a significant role.  The
symbol $\mathbb{E}\{\ldots\}$ denotes ensemble average, so that
\begin{equation}
\label{msd}
\mathbb{E}\left\{{\bf B}^2(t)\right\} = 2 d D t \,,
\end{equation}
$D$ being the ensemble-average diffusion coefficient.
Consequently, the random variable $u$ is defined as the ratio of the
realization-dependent diffusion coefficient, calculated as the
weighted time-average along a single trajectory, and the ensemble
average diffusion coefficient. Clearly, $\mathbb{E}\left\{u\right\} \equiv 1$.

Further on, we analyze here a useful  measure of sample-to-sample
fluctuations - the distribution function $P(\omega)$ of the random
variable
\begin{equation}
\label{omega}
\omega = \frac{u_1}{u_1 + u_2},
\end{equation}
where $u_1$ and $u_2$ are two identical independent random variables
with the distribution $P(u)$. Hence, the distribution $P(\omega)$
probes the likelihood of the event that the diffusion coefficients
drawn from two different trajectories are equal to each other.

Finally, we discuss the effect of disorder on the distributions $P(u)$
and $P(\omega)$ for 1D BM in random media.  We consider two different
models of diffusion in 1D random environments - diffusion
in presence of a random quenched potential with a finite correlation
length, as exemplified here by the Slutsky-Kardar-Mirny model
\cite{slutsky}, and diffusion in a random forcing landscape - the
so-called Sinai model \cite{Sinai1982}. The former is appropriate
for diffusion of proteins on DNA, which is affected by the base-pair
reading interaction and thus is sequence dependent, while the latter
describes, for example, the dynamics of the helix-coil boundary in a melting
heteropolymer \cite{pgg}. Note that in the former case, at
sufficiently large times, one observes a diffusive-type motion with
$B_t^2 \sim t$, while in the latter case dynamics is strongly
\textit{anomalous} so that $B_t$ is logarithmically confined, $B_t^2
\sim \ln^4t$.

The paper is outlined as follows: In Section \ref{a} we recall some common    
fitting procedures used to calculate the diffusion coefficient from single particle tracking data. In Section \ref{b}
we focus on the maximum likelihood estimator and, generalizing 
the approach developed in Ref.~\cite{boyer} for 1D systems,  
obtain new results for the moment generating function $\Phi(\sigma)$  and the
probability density function $P(u)$ 
of the ML estimator for arbitrary 
spatial dimension $d$. In that Section we also obtain the asymptotical behavior of the
probability distribution function $P(u)$, as well as its kurtosis and skewness. 
Next, 
in Section \ref{c}
we focus on the probability distribution function
of the random variable $\omega$ - a novel 
statistical 
diagnostics of the broadness of the parental distribution $P(u)$
which probes the likelihood of the event that two estimates of the diffusion coefficient
drawn from two different trajectories are the same. Further on, Section \ref{d}
presents a comparison 
of the commonly used least squares estimator and the maximum likelihood estimator. We show that the latter 
outperforms the former in any spatial dimension $d$ 
producing a lower variance and the most probable value  
being closer to the ensemble average value. Next, in Section \ref{e} we focus on Brownian motion
in presence of disorder. As exemplified by two 
models 
of dynamics in systems with quenched disorder - Sinai diffusion (random force)
and Slutsky-Kardar-Mirny model (random potential), 
disorder substantially enhances
the importance of sample-to-sample fluctuations. We show that the observation 
of values of the diffusion coefficient 
significantly lower than the ensemble 
average becomes more probable. We show, as well, that as 
the strength of disorder is increased, the distribution $P(\omega)$ undergoes a surprising shape-reversal
transition from a bell-shaped unimodal to a bimodal form with a local minimum at $\omega = 1/2$.
Finally, we conclude in Section \ref{f} with a brief recapitulation of our results and some outline 
of our further research.

\section{Fits for the diffusion coefficient of a single trajectory}
\label{a}

To set up the scene, we first briefly recall several fitting
procedures commonly used to calculate the diffusion coefficient from
the SPT data. More detailed discussion can be found
in Refs.~\cite{saxton,goulian,saxton2,boyer}.
We focus here on estimators which yield a first
 power of $D$. Non-linear estimators, e.g. a mean maximal
 excursion method \cite{ralf} which has been
 used to study anomalous diffusion and
 produces $\sqrt{D}$, will be analyzed elsewhere.

One of the simplest
methods consists in calculating a least squares estimate
based on the minimization of the integral
\begin{equation}
\int^{T}_{0} dt \left({\bf B}^2(t) - l(t)\right)^2,
\end{equation}
where the diffusion law $l(t)$ is taken either as a linear, $l(t) = 2
d D_l t$, or an affine function, $l(t) = 2 d D_a t + b_a$. In
particular, for the linear case the least squares minimization yields
the following linear-least-squares estimator:
\begin{equation}
\label{alpha}
u_{ls} =  \frac{A}{T} \int^T_0 dt \, t \, {\bf B}^2(t),
\end{equation}
where $A$ is the normalization factor, $A = 3/2 d D T^2$, conveniently
chosen so that $\mathbb{E}\left\{u_{ls}\right\} \equiv 1$.

A second, more sophisticated, approach is based on
\begin{equation}
\label{delta}
\delta^2_T(t) = \frac{1}{T - t} \int^{T - t}_0 \, dt' \, \left({\bf B}(t' + t) - {\bf B}(t')\right)^2 \,,
\end{equation}
which is the temporal moving
average over a sufficiently long trajectory ${\bf B}(t)$ produced by the
underlying process of duration $T \gg t$.  The diffusion coefficient
is then extracted from fits of $\delta^2_T(t)$, or from a related
least squares estimator, which is given by the following functional of
the trajectory
\begin{equation}
\label{deltaT}
u_{\delta} = \frac{A}{T} \int^T_0 \, dt \, t \, \delta^2_T(t),
\end{equation}
where $A$ is the same normalization constant as in Eq.~(\ref{alpha}).
Note that the random variable $u_{\delta}$ is again conveniently
normalized so that $\mathbb{E}\left\{u_{\delta}\right\} \equiv 1$, which enables a direct comparison
of the respective distributions of different estimators.
As shown in
Ref.~\cite{boyer}, $u_{\delta}$  only provides a 
slightly better estimate of
$D$ than $u_{ls}$.
A conceptually different fitting procedure has been discussed in
Ref.~\cite{boyer} which amounts to maximizing the unconditional
probability of observing the whole trajectory ${\bf B}(t)$, assuming
that it is drawn from a Brownian process with mean-square displacement
$2 d D t$ (see Eq.~(\ref{msd})).  This is the maximum likelihood estimate which
takes the value of $D$ that maximizes the likelihood of ${\bf B}(t)$,
defined as:
\begin{equation}
L = \prod_{t = 0}^T \left(4 \pi dD t\right)^{-d/2} \exp\left( - \frac{{\bf B}^2(t)}{4 dD t}\right)
\end{equation}
where the trajectory ${\bf B}(t)$ is appropriately discretized. Differentiating the
logarithm of $L$ with respect to $D$ and setting $d \ln L/d D =
0$, one finds the maximum likelihood estimate of $D$, which upon a
proper normalization is defined by Eq.~(\ref{u}).

Below we will derive
the distribution function $P(u)$ of the ML estimator
$u$ and compare it against numerical results for the distribution
function of the LS estimator $u_{\delta}$ for $d = 1,2,3$.

\section{Distribution of the ML estimator}
\label{b}

\subsection{The moment generating function}

Let $\Phi(\sigma)$ denote the moment generating function of the random
variable $u$ defined in Eq.~(\ref{u}),
\begin{equation}
\Phi(\sigma) = \mathbb{E}\left\{e^{- \sigma u}\right\}.
\end{equation}
The squared distance from the origin, of $d$-dimensional BM at time
$t$ for a given realization, decomposes into the sum
\begin{equation}
{\bf B}^2(t) = \sum_{i = 1}^d B_i^2(t),
\end{equation}
$B_i(t)$ being realizations of trajectories of independent
1D BMs (for each spatial direction). Thus, $\Phi(\sigma)$
factorizes
\begin{equation}
\Phi(\sigma) = G^d(\sigma),
\end{equation}
where
\begin{equation}
G(\sigma) = \mathbb{E}\left\{\exp\left( - \frac{\sigma}{2 d D T} \int^T_0 d\tau \frac{B_i^2(\tau)}{\tau} \right)\right\}.
\end{equation}

Here, in order to calculate $G(\sigma)$, we follow the strategy of
Ref.~\cite{boyer} and introduce an auxiliary functional:
\begin{equation}
\label{fc}
\Psi(x,t) = \mathbb{E}^x_t\left\{ \exp\left( - \frac{\sigma}{2 d D T} \int^T_t d\tau \frac{B_i^2(\tau)}{\tau}    \right)\right\}
\end{equation}
where the expectation is for a BM starting at $x$ at time $t$.  We
derive a Feynman-Kac type formula for $\Psi(x,t)$ considering how the
functional in Eq.~(\ref{fc}) evolves in the time interval
$(t,t+dt)$. During this interval the BM moves from $x$ to $x +
dB_i(t)$, where $dB_i(t)$ is an infinitesimal Brownian increment such
that $\mathbb{E}_{dB}\{dB_i(t)\} = 0$ and $\mathbb{E}_{dB}\{dB^2_i(t)\} = 2 D dt$, where
$\mathbb{E}_{dB}$ denotes now averaging with respect to the increment
$dB_i(t)$. For such an evolution we have to order $dt$:
\begin{eqnarray}
\Psi(x,t) &=& \mathbb{E}\Big\{ \left(1 - \frac{\sigma x^2}{2 d D T t} dt\right) \nonumber\\
&\times&
\mathbb{E}^{x+dB_i(t)}_{t+dt}\left\{ \exp\left( - \frac{\sigma}{2 d D T} \int^T_{t+dt} d\tau \frac{B_i^2(\tau)}{\tau}    \right)\right\}  \Big\} \nonumber\\
&=& \mathbb{E}\left\{ \Psi(x + dB_i(t),t + dt) \left(1 - \frac{\sigma x^2}{2 d D T t} dt\right)\right\}
\end{eqnarray}
Expanding the right-hand-side of the latter equation to second order
in $dB_i(t)$, linear order in $dt$ and performing averaging, we find
eventually the following Schr\"odinger equation:
\begin{equation}
\frac{\partial \Psi(x,t)}{\partial t} = - D \frac{\partial^2 \Psi(x,t)}{\partial x^2} + \frac{\sigma x^2}{2 d D T t} \Psi(x,t).
\end{equation}
The solution of this equation has been obtained in Ref.~\cite{boyer}
and gives
\begin{equation}
G(\sigma) = \Psi(0,0) = \left(I_0\left(\sqrt{8 \sigma/d}\right)\right)^{-1/2},
\end{equation}
where $I_0(.)$ is the modified Bessel function \cite{abramowitz}.
Consequently, we find the following general result
\begin{equation}
\label{t}
\Phi(\sigma) = \left(I_0\left(\sqrt{8 \sigma/d}\right)\right)^{-d/2}.
\end{equation}
Note that $\Phi(\sigma)$ is independent of $T$ and $D$, as
it should in virtue of the scaling properties of the BM.

\subsection{The distribution function}

We turn next to the analysis of the distribution of the ML estimator defined in Eq.~(\ref{u}).
First of all, we calculate several
first moments of $u$ by merely differentiating the result in Eq.~(\ref{t}):
\begin{eqnarray}
\mathbb{E}\left\{u^2\right\}  &=& 1 + \frac{1}{d}\nonumber\\
\mathbb{E}\left\{u^3\right\}  &=& 1 + \frac{3}{d} + \frac{8}{3 d^2}\nonumber\\
\mathbb{E}\left\{u^4\right\}  &=& 1 + \frac{6}{d} + \frac{41}{3 d^2} + \frac{11}{d^3}
\end{eqnarray}
Consequently, one may expect that all moments tend to $1$ as $d \to
\infty$, so that $P(u) \to \delta(u - 1)$.  For fixed $d$, the
variance $\mathbb{E}\left\{u^2\right\} - \mathbb{E}\left\{u\right\}^2 \equiv 1/d$, the
coefficient of asymmetry $\gamma_a \equiv 8/3\sqrt{d}$ and the
kurtosis $\gamma_e \equiv 11/d$. All these characteristics vanish when
$d \to \infty$.

\begin{figure}[ht]
  \centerline{\includegraphics*[width=0.45\textwidth]{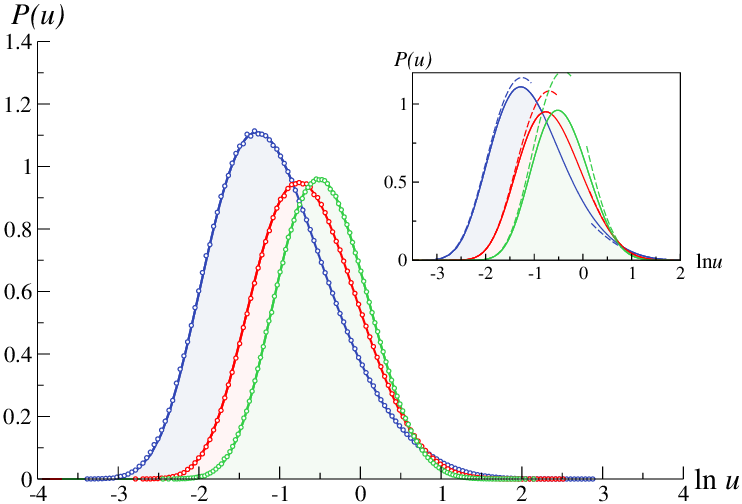}}
  \caption{(Color online) The distribution $P(u)$ in Eq.~(\ref{integral}) 
  for ML estimates of BM: the blue solid
    line (left) corresponds to $d=1$, the red line (middle) to $d=2$ and the 
    green line (right) to $d=3$. The
    open circles present the results of
    numerical simulations.  In the inset the dashed lines correspond
    to the small-$u$ and large-$u$ asymptotics in Eqs.~(\ref{small}) and
    (\ref{large}).}
\label{fig1}
\end{figure}

Note next that since $I_0(\sqrt{8 \sigma/d}) = J_0(i \sqrt{8
  \sigma/d})$, the poles of $\Psi(\sigma)$ are located at $\sigma = -
d \gamma_k^2/8$, where $\gamma_k$ is the $k$th zero of the Bessel
function $J_0(.)$ \cite{abramowitz}. Consequently, for even $d$, we
can straightforwardly find $P(u)$ in form of an infinite series in the
zeros of the Bessel function $J_0(.)$.  For $d = 2$, $\Phi(\sigma)$
has only simple poles so that the expansion theorem gives
\begin{eqnarray}
\label{2D}
P(u) &=&  \frac{1}{2} \sum_{k = 1}^{\infty} \frac{\gamma_k}{J_1(\gamma_k)} \, \exp\left( - \frac{\gamma_k^2}{4} u\right).
\end{eqnarray}
For $d=4$ and $d = 6$ the standard residue calculus yields
\begin{equation}
\label{4D}
P(u) = \sum_{k = 1}^{\infty} \frac{\gamma_k \left(\gamma_k \, J_1(\gamma_k) \, u - J_2(\gamma_k)\right)}{J_1^3(\gamma_k)} \, \exp\left( - \frac{\gamma_k^2}{2} u\right)
\end{equation}
and
\begin{equation}
\label{6D}
P(u) = \frac{3}{16} \sum_{k=1}^{\infty} \frac{\alpha_k}{\gamma_k J_1^5(\gamma_k)} \exp\left( - \frac{3\gamma_k^2}{4} u\right),
\end{equation}
where
\begin{equation}
\alpha_k = 9 \gamma_k^4 J_1^2(\gamma_k) u^2 - 36 \gamma_k^2 J_1^2(\gamma_k) u + 4 J_2^2(\gamma_k) (\gamma_k^2 - 8).
\end{equation}
Similar results can be readily obtained for greater even $d$.

For arbitrary $d$, including odd values, the distribution $P(u)$ is
defined by inverting the Laplace transform and is given by the
following integral:
\begin{equation}
\label{integral}
P(u) = \frac{1}{\pi} \int^{\infty}_0 \frac{\cos\left(u y - d \phi/2\right) \, dy}{\left({\rm ber}^2(\sqrt{8 y/d}) +
{\rm bei}^2(\sqrt{8 y/d})\right)^{d/4}}
\end{equation}
where the phase $\phi$ is given by
\begin{equation}
\label{phi}
\phi = \sum_{k=1}^{\infty} {\rm Arcsin}\left(\frac{8 y}{\sqrt{d^2 \gamma_k^4 + 64 y^2}}\right) =  {\rm Arctan}\left(\frac{{\rm bei}(\sqrt{\frac{8 y}{d}})}{{\rm ber}(\sqrt{\frac{8 y}{d}})}\right)
\end{equation}
${\rm ber}(x)$ and ${\rm bei}(x)$ being the $0$th order Kelvin
functions \cite{abramowitz}.

Finally, we consider the small-$u$ and large-$u$ asymptotic behavior
of the probability density function $P(u)$.  To extract the small-$u$
asymptotic behavior of $P(u)$ we consider the large-$\sigma$ form of
$\Phi(\sigma)$. From Eq.~(\ref{t}) we get
\begin{equation}
\Phi(\sigma) \sim \left(2 \pi\right)^{d/4} \left(\frac{8 \sigma}{d}\right)^{d/8} \exp\left(- \sqrt{2 d \sigma}\right)
\end{equation}
as $\sigma \to \infty$.  Consequently, we find the following strongly
non-analytical behavior:
\begin{equation}
\label{small}
P(u) \sim (4 \pi)^{d/4} \sqrt{\frac{d}{2 \pi} } \exp\left(- \frac{d}{2 u}\right) \frac{1}{u^{1 + \mu}}, \, \mu = \frac{d + 2}{4} \,.
\end{equation}
The large-$u$ behavior of the distribution $P(u)$ is defined by the
behavior of the moment generating function $\Phi(\sigma)$ in the
vicinity of $\sigma^* = - d \gamma_1^2/8$,
\begin{equation}
\Phi(\sigma) \sim \left(\frac{d \gamma_1}{4 J_1(\gamma_1)}\right)^{d/2} \, \frac{1}{\left(\sigma + d \gamma_1^2/8\right)^{d/2}} \,.
\end{equation}
Consequently, we find that for $u \to \infty$, $P(u)$ decays as
\begin{equation}
\label{large}
P(u) \sim \frac{1}{\Gamma(d/2)} \, \left(\frac{d \gamma_1}{4 J_1(\gamma_1)}\right)^{d/2} \, u^{d/2 - 1} \, \exp\left(- \frac{d \gamma_1^2}{8} u\right) \,.
\end{equation}
This behavior is, of course, consistent with the series expansions in Eqs.~
(\ref{2D}), (\ref{4D}) and (\ref{6D}). Our results on the distribution
$P(u)$ are summarized in Fig.~\ref{fig1}.

\begin{figure}[ht]
  \centerline{\includegraphics*[width=0.45\textwidth]{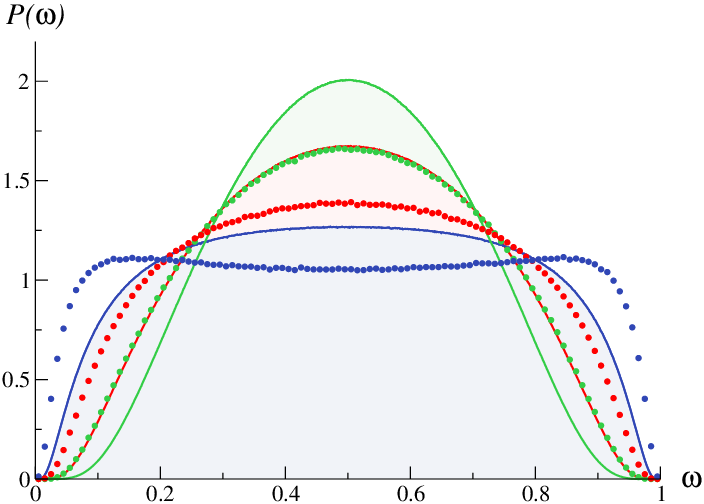}}
  \caption{(Color online) Distribution $P(\omega)$ for standard BM obtained 
  from Eqs.~ (\ref{integral}) and (\ref{pomega}). $P(\omega=1/2)$ 
  increases with the dimension $d$. The blue (lower) 
  solid line corresponds
    to $d=1$, the red (middle) line to $d = 2$ and the green (upper) line 
    to $d=3$.  Open circles
    with the same color code (and relative positions) present the results of numerical
    simulations for the LS estimator $u_{\delta}$ defined in Eq.~(\ref{deltaT}).
  Note that an apparent coincidence 
of the results for the distributions $P(\omega)$ for the ML estimator in 2D 
and that for 
the LS estimator $u_{\delta}$ in 3D is 
accidental. It just signifies that the former outperforms the latter. 
  }
  \label{2DP(w)}
\end{figure}

\section{The distribution of the random variable $\omega$}
\label{c}

Suppose next that we have two different independent realizations of
BM trajectories, ${\bf B}_1(t)$ and ${\bf B}_2(t)$ which
we use to generate to independent random variables $u_1$ and
$u_2$. A natural question arising about their suitability as estimators is how  likely is it that they will have the same value? Of course the distributions and thus moments of these two random variables are the same, however a measure of their relative dispersion can be deduced by
studying the distribution function $P(\omega)$ of the random variable
$\omega$ \cite{carlos,carlos1}, defined in Eq.~(\ref{omega}). This distribution
is given explicitly by \cite{gleb}
\begin{equation}
\label{pomega}
P(\omega) = \frac{1}{(1 - \omega)^2} \int^{\infty}_0 u \, du \, P(u) \, P\left(\frac{\omega}{1 - \omega} u\right)
\end{equation}
and hence, it suffices to know $P(u)$ in order to
determine $P(\omega)$.

For $d = 2$, (and, in fact, for
any other even $d$), $P(\omega)$ can be evaluated exactly. Plugging
Eq.~(\ref{2D}) into (\ref{pomega}), we get:
\begin{equation}
P(\omega) = 4 \sum_{k,l = 1}^{\infty} \frac{\gamma_k}{J_1(\gamma_k)} \frac{\gamma_l}{J_1(\gamma_l)} \frac{1}{\left((1 - \omega) \gamma_k^2 + \omega \gamma_l^2 \right)^2}
\end{equation}
Performing the sum over $l$,
%Further on, we make use of an identity
%\begin{equation}
%\frac{1}{\left((1 - \omega) \gamma_k^2 + \omega \gamma_l^2 \right)^2} = \frac{1}{\gamma_k^2} %\frac{d}{d\omega} \frac{1}{\gamma_l^2 + ((1- \omega)/\omega) \gamma_k^2}
%\end{equation}
%and get
%\begin{eqnarray}
%P(\omega) = 4 \frac{d}{d \omega} \sum_{k,l = 1}^{\infty} \frac{\gamma_l}{\gamma_k J_1(\gamma_k) %J_1(\gamma_l)} \frac{1}{\gamma_l^2 + ((1- \omega)/\omega) \gamma_k^2}.
%\end{eqnarray}
%Using next the definition
%\begin{eqnarray}
%\Phi(\sigma) &=& \int^{\infty}_0 du \exp\left(- \sigma u\right) \, P(u) = \nonumber\\
%&=& 2 \sum_{l = 1} \frac{\gamma_l}{J_1(\gamma_l)} \frac{1}{\gamma_l^2 + 4 \sigma} = \frac{1}{I_0\left(2 %\sqrt{\sigma}\right)},
%\end{eqnarray}
we arrive at the following result for the distribution $P(\omega)$ in
2D systems
\begin{equation}
\label{2dfinal}
P(\omega) = 2 \frac{{\rm d}}{{\rm d} \omega} \sum_{k = 1}^{\infty} \frac{1}{\gamma_k J_1(\gamma_k) I_0(\gamma_k \sqrt{(1-\omega)/\omega})} \,.
\end{equation}
Numerically obtained distributions $P(\omega)$ for $d=1, 2$ and $d=3$
dimensional systems are presented in Fig.~\ref{2DP(w)}. Notice that in
all dimensions $\omega = 1/2$ is the most probable value of the
distribution $P(\omega)$ so that most probably $u_1 =
u_2$. Nevertheless,  the distributions are rather broad which signifies
that sample-to-sample fluctuations are rather important.

\section{Comparison of the LS and ML estimators.}
\label{d}

We will now show that the ML estimator defined in Eq.~(\ref{u})
substantially outperforms the MS estimator as defined in Eq.~(\ref{deltaT}) in any
spatial dimension $d$. This is a very surprising result as one would intuitively expect, and it is often stated in the literature, that using the process $\delta_T$ has the effect of a reducing the fluctuations of the estimate of $D$ because the process is partially averaged in time.

\begin{figure}[ht]
  \centerline{\includegraphics*[width=0.45\textwidth]{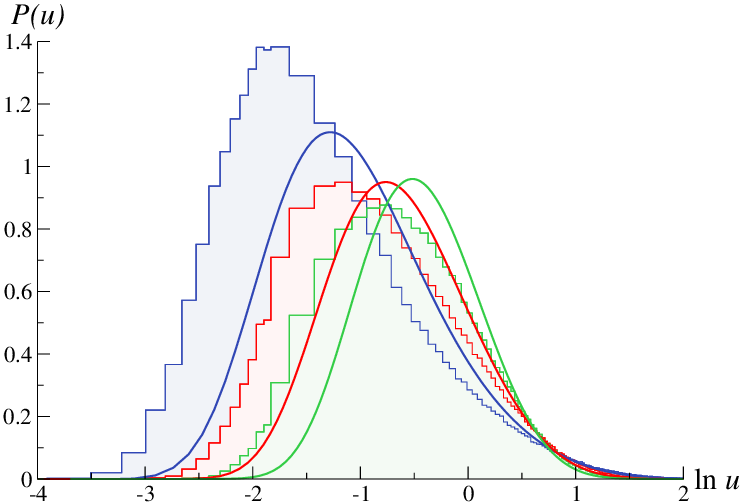}}
  \caption{(Color online) Comparison of $P(u)$ in Eq.~(\ref{integral}) 
  (solid lines) and the results of the numerical simulations for 
  the distribution $P(u_{\delta})$ (histograms). From
  left to right: $d=1$, 2 and 3.}
  \label{comp}
\end{figure}

To demonstrate this, we present in Fig.~\ref{comp}
a comparison of the analytical results for $P(u)$ of the ML estimator with the corresponding
distributions of the LS estimator  for $u_\delta$ obtained
numerically.

Indeed, we find that the variance of the distribution $P(u_{\delta})$
equals $1.38$, $0.66$ and $0.44$ for $d=1, 2$ and $3$,
respectively. The distribution of the ML estimator appears to be
substantially narrower so that the variance is significantly lower,
$1$, $1/2$ and $1/3$. Moreover, the most probable values of $u$ are
closer to the ensemble average value $\mathbb{E}\{u\} \equiv 1$ than the most
probable values of $P(u_{\delta})$ to $\mathbb{E}\{u_{\delta}\} \equiv 1$: we
observe that the distribution $P(u_{\delta})$ attains its maximal
values at $u_{\delta} \approx 0.15$, $0.33$ and $0.47$ for $d=1, 2$
and $3$, respectively, while the corresponding maxima of the
distribution $P(u)$ are located at $u \approx 0.28, 0.47$ and $0.6$.
Last but not least, the distribution $P(\omega_{\delta})$ appears to
be significantly broader than $P(\omega)$, as revealed by
Fig.~\ref{2DP(w)}. The worst performance of the LS estimator
$u_{\delta}$ is in 1D systems in which the distribution
$P(\omega_{\delta})$ has a bimodal  $M$-shape with a
local \textit{minimum} at $\omega_{\delta} = 1/2$, and maxima (most likely values)
around $0.1$ and $0.9$. 
This means that the values $u_1$ and $u_2$ drawn from two different trajectories
will most probably be different by an order of magnitude!

\section{1D Brownian motion in presence of disorder.}
\label{e}

In this final section we address the question of how the distribution
$P(u)$ of the ML estimator of a single trajectory diffusion
coefficient will change in presence of quenched disorder.  We will consider two
different models of BM in random 1D environments - diffusion in
presence of a random correlated potential and diffusion in presence of
a random force.

\begin{figure}[ht]
  \includegraphics[width=0.45\textwidth]{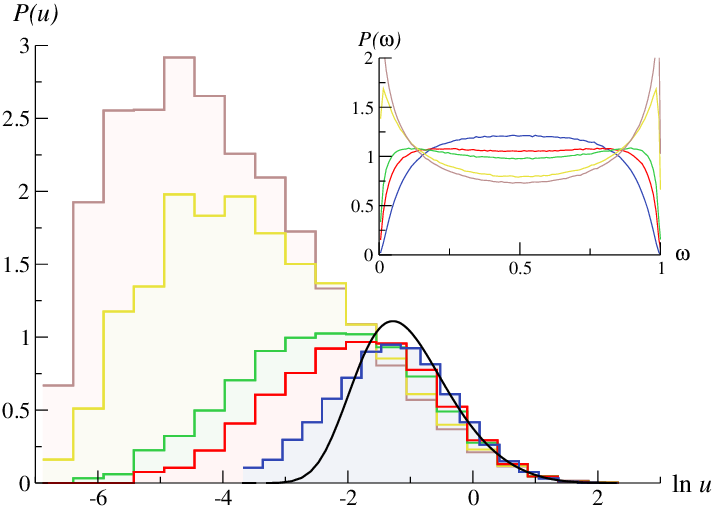}
  \caption{(Color online) Distribution $P(u)$ for a particle diffusing in a 
   random
    energy landscape with variance $\sigma^2$, correlation length
    $\xi_c=20$ and various disorder strengths $\epsilon = \beta
    \sigma$. From right to left: blue histogram corresponds to 
    $\epsilon=0.5$, red - to
    $\epsilon=0.8$, green - to $\epsilon=1$, yellow - to $\epsilon=2$
    and brown - to $\epsilon=3$.  The walk duration is $T=10^5$, $t_0=0$ and
    averages are taken over $12,000$ walks occurring in independent
    landscapes. For comparison we present  the distribution (solid black curve) for
    standard 1D BM ($\varepsilon \equiv 0$). The corresponding distributions 
    $P(\omega)$ are shown in the inset 
    ($P(\omega=1/2)$ decreases with increasing $\epsilon$).}
  \label{figPuMirny}
\end{figure}

\subsection{Diffusion in presence of a random potential}
\label{z}
First we consider a BM in a 1D inhomogeneous
energy landscape, where disorder is correlated over a finite length
$\xi_c$.  This model gives a simple description of diffusion of a
protein along a DNA sequence, for instance, where the particle
interacts with several neighboring base pairs at a time
\cite{slutsky}.  The total binding energy of the protein is assumed to
be a random variable.  When the particle hops one neighboring base
further to the right or to the left, its new energy is highly
correlated to the value it had before the jump. Slutsky {\it et al.}
\cite{slutsky} modeled this process as a point-like particle diffusing
on a 1D lattice of unit spacing with random site energies $\{U_i\}$, whose
distribution is Gaussian with zero mean, variance $\sigma^2$ and is
correlated in space as
$\langle(U_i-U_j)^2\rangle=2\sigma^2[1-\exp(-|i-j|/\xi_c)]$. At each
time step, the particle located at some site $i$ jumps to the left or
to the right with probabilities $p_i \propto \exp[\beta(U_i-U_{i-1})]$
and $q_i \propto \exp[\beta(U_i-U_{i+1})]$, respectively, where
$p_i+q_i=1$.  Diffusion is asymptotically normal for any disorder
strength $\epsilon = \beta \sigma$.  Nevertheless, the particle can be
trapped in local energy minima for long periods of time. During an
extended intermediate time regime, it is observed that first passage
properties fluctuate widely from one sample to another \cite{slutsky}.

Our numerical simulations reveal that disorder has a dramatic effect
on the distributions $P(u)$ and $P(\omega)$.  As shown by
Fig.~\ref{figPuMirny}, the distribution $P(u)$ broadens significantly
in the small $u$ regime: very small values of the time-average diffusion
constant (compared to the thermal and disorder average) become
increasingly more probable as the disorder strength increases.
However, the right tail of $P(u)$ is much less affected.  Similarly, two
independent measurements are likely to differ significantly, even in
moderately disordered media (see inset of Fig.~\ref{figPuMirny}).
When $\epsilon \approx 0.8$, the distribution $P(\omega)$ undergoes
a continuous shape reversal transition - from a unimodal bell-shaped
form to a bimodal $M$-shape one with the minimum at
$\omega = 1/2$ and two maxima approaching $0$ and $1$ at larger
disorder strengths. Unfortunately, it does not 
seem possible to obtain this critical value analytically. Even for 
the case of a pure Brownian motion considered in Ref.~\cite{carlos}, 
such an analysis appears to be extremely difficult.

Therefore,  for  $\epsilon  >  0.8$  sample-to-sample  fluctuations
becomes  essential   and  it  is   most  likely  that   the  diffusion
coefficients drawn from two different trajectories will be different.

\begin{figure}[ht]
  \includegraphics[width=0.45\textwidth]{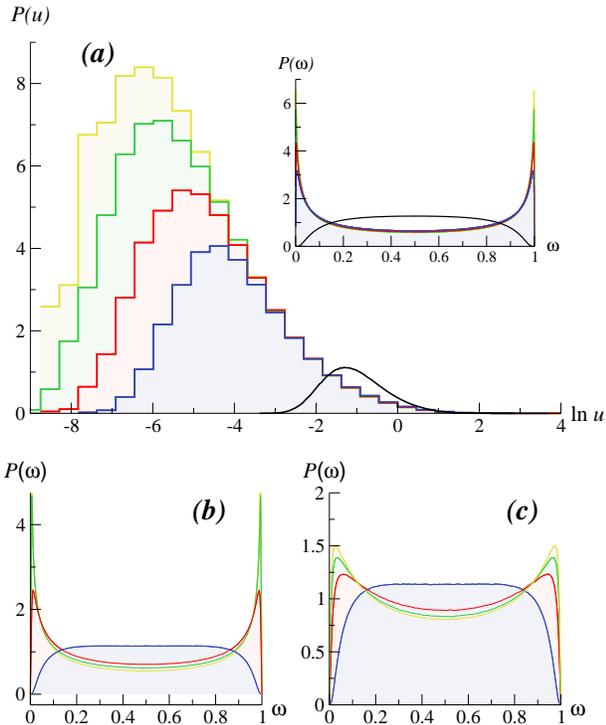}
  \caption{(Color online) In panel {\it (a)}, distribution $P(u)$ for Sinai 
  diffusion
    and different strengths of the disorder $\epsilon$. From
    right to left: blue histogram
    corresponds to $\epsilon=0.05$, red - to $\epsilon=0.1$, green -
    to $\epsilon=0.2$ and yellow - to $\epsilon=0.3$. The solid black
    curve depicts the distribution $P(u)$ for 1D BM
    ($\epsilon \equiv 0$), and the corresponding distributions
    $P(\omega)$ are shown in the inset with the same color code 
    (for $\omega$ close to 0 or 1, $P(\omega)$ increases with $\epsilon$). In
    panel {\it (b)}, distribution $P(\omega)$ for Sinai diffusion with
    $\epsilon=0.1$, the integration time is set to $T=10^2$ and  $t_0$ is
    varied. $P(\omega=1/2)$ decreases with increasing $t_0$: the
    different curves corresponds to
    (from darker gray to lighter gray): $t_0=5$ (blue), $t_0=5 \times 10^2$ (red),
    $t_0=5 \times 10^3$ (green) and $t_0=5\times 10^5$ (yellow). 
    In panel {\it (c)},
    the lag time is set to $t_0=5$ and the integration time $T$
    is varied. $P(\omega=1/2)$ decreases with increasing $T$:
      $10^2$ (blue), $5 \times 10^4$ (red), $1,5 \times 10^5$ (green) and $2,5\times 10^5$
    (yellow).}
  \label{fig:sinai}
\end{figure}

\subsection{Diffusion in presence of a random force}

We discuss now the effect of disorder on the distributions $P(u)$ and
$P(\omega)$ for 1D BM in presence of a quenched uncorrelated random
force - the so-called Sinai diffusion \cite{Sinai1982}.  In this
model, one considers a random walk on a 1D infinite lattice and site $i$
dependent hopping probabilities: $p_i =
\frac{1}{2} - \varepsilon_i$ for hopping from $i$ to the site $i+1$
and $q_i = \frac{1}{2} + \varepsilon_i$ for hopping to the site
$i-1$. Here, $\varepsilon_i$ are independent, uncorrelated,
identically distributed random variables with distribution
$P(\varepsilon_i) = \frac{1}{2}\delta(\varepsilon_i-\varepsilon) +
\frac{1}{2}\delta(\varepsilon_i+\varepsilon)$ and the strength of
disorder $\varepsilon$ is bounded away from $0$ and $1$. It is
well-known that in the large-$t$ limit the model produces an
anomalously slow sub-diffusion $\langle x^2(t)\rangle \sim \ln^4(t)$, where the angle brackets denote averaging with respect to different realizations of disorder.
At shorter times, however, one observes an extended stage with a
transient behavior which is substantially different from the
asymptotic one.  As a consequence, the statistics of $u$, defined in equation \eqref{u},
and $\omega$ will depend not only on the integration time $T$ but also
on the lag time $t_0$ from which a single trajectory is analyzed.

We have numerically computed the distributions $P(u)$ and $P(\omega)$.
In Fig.~\ref{fig:sinai}$a$ we present the dependence of the $u$ and
$\omega$ statistics on the strength of the disorder $\varepsilon$.  As
in the previous disordered potential case, we find that the maximum of
$P(u)$ shifts toward zero as the disorder gets stronger.  For
comparison, the solid black line in Fig.~\ref{fig:sinai}$a$ represents
$P(u)$ observed for standard BM, $\varepsilon \equiv 0$.
Moreover, the stronger the disorder is, the broader the distribution
$P(u)$ becomes, yielding more peaked maxima in $P(\omega)$ (see the
inset of Fig.~\ref{fig:sinai}$a$).  We also note that $P(\omega)$ has
a bimodal $M$-shaped form even for the weakest disorder
$\varepsilon=0.005$ that we have considered, suggesting that the
zero-disorder limit is non-analytic compared to the continuous
transition observed for diffusion in the random energy landscape of the previous section.

In Fig.~\ref{fig:sinai}$b$ and Fig.~\ref{fig:sinai}$c$, we show the
statistics of $\omega$ for different values of $t_0$ and
$T$. Increasing $t_0$ (or $T$) we observe that $P(\omega)$ changes
from an almost uniform form, for which any relation between
$u_1$ and $u_2$ is equally probable, to a bimodal $M$-shaped
distribution, which signifies that in this regime
two ML estimates $u_1$ and $u_2$ will most likely have different values.  
Bimodality is a property of the Sinai regime
\cite{carlos}, as also noticed earlier in Ref.~\cite{redner}, and thus
it shows up only at sufficiently long times when the trajectories
follow the asymptotic ultra-slow diffusion.  The distribution
$P(\omega)$ is remarkably sensitive to the characteristic aging of
Sinai diffusion.

As a final observation, one may also study the statistics of $u$ for
trajectories evolving in the same random force field. In this case
(not shown here) one gets a narrow distribution for of $u$ and a
unimodal $P(\omega)$ that converges to a delta singularity at
$\omega=1/2$ when the disorder becomes infinite.  This is due to the
fact, known as Golosov phenomenon,  that two trajectories in the same disorder will move together
\cite{Golosov1983}.

\section{Discussion}
\label{f}

We have analyzed the reliability of the ML estimator for the diffusion
constant of standard Brownian motion and shown its superiority over
the more commonly used LS estimator in a number of important aspects,
notably the variance of the estimator, the proximity of the most
probable value to the true mean value and the distribution of the
random variable $\omega$ which is a measure of the extent to which two
estimations of $D$ vary. Going beyond the important test case of pure
Brownian motion we have also analyzed the effect of quenched disorder,
modeling fluctuations of the local energy landscape and forces. As one
may have intuitively expected, the presence of short range disorder
tends to broaden the distribution of the so measured value of $D$, as
it presents an additional source of fluctuation. However in the Sinai
model, in the same realization of the force field, trajectories are
disorder dominated and are almost independent of the thermal noise,
leading to highly peaked distributions of $D$.  Analytically
understanding the distribution of $D$ in the presence of disorder
presents an interesting mathematical challenge which will involve
analysis of the corresponding Schr\"odinger equation with a random
drift term. Further interesting questions remain to be addressed, in
particular can one use the two time correlation function of measured
trajectories to obtain better estimators ? Single particle tracking
technology will undoubtedly further improve in the coming years and
many interesting mathematical, statistical and physical challenges
will arise in the ultimate goal of getting the most out of the
trajectories so obtained.

\begin{acknowledgments}
  DSD, CMM and GO gratefully acknowledge support from the ESF and
  hospitality of NORDITA where this work has been initiated during
  their stay within the framework of the "Non-equilibrium Statistical
  Mechanics" program.
\end{acknowledgments}

\end{document}